\documentclass[aps,pre,twocolumn,groupaddress]{revtex4-1}
\usepackage{color}
\usepackage{graphicx}
\usepackage{amsmath}
\usepackage{amsfonts}
\usepackage{amssymb}

\begin{document}
\title{Conundrum of weak noise limit for diffusion in a tilted periodic potential}
%
\author{J. Spiechowicz}
\affiliation{Institute of Physics, University of Silesia, 41-500 Chorz{\'o}w, Poland}
\author{J. {\L}uczka}
\affiliation{Institute of Physics, University of Silesia, 41-500 Chorz{\'o}w, Poland}
\email{jakub.spiechowicz@us.edu.pl}
\begin{abstract}
The weak noise limit of dissipative dynamical systems is often the most fascinating one. In such a case fluctuations can interact with a rich complexity frequently hidden in deterministic systems to give rise of completely new phenomena that are absent for both noiseless and strong fluctuations regimes. Unfortunately, this limit is also notoriously hard to approach analytically or numerically. We reinvestigate in this context the paradigmatic model of nonequlibrium statistical physics consisting of inertial Brownian particle diffusing in a tilted periodic potential by exploiting the state of the art computer simulations of unprecedented time scale. In contrast to the previous results on this long standing problem we draw an inference that in the parameter regime for which the particle velocity is bistable  the lifetime of ballistic diffusion diverges to infinity 
when thermal noise intensity tends to zero, i.e. an everlasting ballistic diffusion  emerges. As a consequence the diffusion coefficient does not reach its stationary constant value.
\end{abstract}
\maketitle
\section{Introduction}
Deterministic nonlinear dynamical systems are of paramount importance due to their vast applications in a broad spectrum of scientific disciplines \cite{strogatz}. Their nonlinear nature allows for a rich complexity which is not present in linear systems. One of the reasons behind it is validity of the \emph{superposition principle} for linear systems which states that the response caused by two or more perturbations is the sum of the reactions that would have been induced by each stimulus individually. Nonlinear setups are known to exhibit unusual behaviour like chaos - in which deterministic nature of a system does not make it predictable \cite{ott}, multistability - when two or more stable states are present in the setup dynamics \cite{pisarchik2014}, limit cycles - existence of asymptotic periodic orbits to which a perturbed system is attracted \cite{strogatz} or solitons - self-reinforcing wave packets \cite{lamb}, to name only a few.

On the other hand, in the last four decades it has been understood that noise can produce qualitative changes in the properties of a deterministic system \cite{sagues2007}. Random fluctuations acting upon a nonlinear setup out of equilibrium may have particularly far-reaching consequences. It is due to the fact that equilibrium is ruled by various Thermodynamic Laws and symmetries, such as e.g. the \emph{detailed balance}, which generally loose their validity out of equilibrium. There are two main ways in which noise can interact with nonlinearity to render counter-intuitive behaviour. First, fluctuations can help multistable system to cross a potential barrier separating different stable states. If, for a fine dose of noise, the random crossing times statistically match a deterministic time scale of the system, a more regular behaviour may emerge e.g. in the form of a (quasi-)periodicity. This is the basic mechanism standing behind phenomena such as stochastic resonance \cite{benzi1981, gammaitoni1998} or coherence resonance \cite{pikovsky1997,lindner2004}. Second, fluctuations can destabilize stationary states existing in the nonlinear dynamics and induce new ones which could correspond to qualitatively and quantitatively different evolution. The latter observation is \emph{modus operandi} of phenomena like noise induced transport \cite{hanggi2009, spiechowicz2014pre}, negative mobility \cite{machura2007,nagel2008,slapik2019,spiechowicz2019njp} or recently discovered fluctuations induced dynamical localization \cite{spiechowicz2017scirep, spiechowicz2019chaos} causing anomalous diffusion \cite{metzler2014} in systems which at the first glance cannot react in this way.

In this context particularly interesting is a limiting situation of \emph{weak noise} interacting with a deterministic nonlinear system. When intensity of fluctuations is too large they usually smear the dynamics too much so that the impact of deterministic evolution is barely visible if not completely irrelevant. In this way a rich complexity often hidden in a noiseless nonlinear system is destroyed. Unfortunately, the weak noise limit is also notoriously hard to approach. Dynamical systems under the influence of fluctuations are often successfully modeled by the Fokker-Plank equation \cite{risken} whose time-independent solutions corresponding to the steady states are the most relevant. Mathematically, the weak noise limit of the time independent solutions of the Fokker-Planck equation is the most problematic one. The reason is that then actually two limits are involved: time tends to infinity, in which the steady state is approached and the weak noise limit that must be carried out after the limit $t \to \infty$.

In this paper we reinvestigate the paradigmatic model of nonequilibrium statistical physics, namely, underdamped Brownian motion in a biased periodic potential to analyze diffusion in the weak noise limit. This setup is isomorphic with many important physical systems \cite{risken} such as Josephson junction \cite{junction}, dipoles rotating in external fields \cite{Coffey}, superionic conductors \cite{Ful1975}, charge density waves \cite{Gru1981} and cold atoms dwelling in optical lattices \cite{denisov2014}, to mention only a few. In view of the above discussion it is not surprising that the asymptotic analytical methods are not yet elaborated for this system and one needs to rely solely on numerical results. Consequently, there are many mutually contradictory results in the literature concerning the conundrum of weak noise limit for diffusion in a tilted periodic potential. Here we exploit the state of the art computer simulations of unprecedented time scale to draw the inference about this long long existent problem.

The paper is organized as follows. In Sec. II we recall the formulation of the model and introduce the dimensionless quantities. 
Next, in Sec. III we describe the basic features of deterministic dynamics. Then, in Sec. IV we introduce the diffusion quantifiers and discuss the state of the art of the weak noise limit. In Sec. V we first comment on the everlasting anomalous diffusion occurring in this system. Next,  we  debate on the weak ergodicity breaking and the relation between the diffusion coefficient and the residence probabilities in two states of the velocity dynamics. In Sec. VI we discuss the finite time measurements of the diffusion coefficient that is important from the experimental point of view. Sec. VII provides a summary and conclusions.

\section{Model}
In this paper we revisit the problem of diffusion in a tilted periodic (also named washboard) potential. We consider a classical Brownian particle of mass $M$, moving in a spatially periodic and \emph{symmetric} potential $U(x) = U(x + L)$ of period $L$ and subjected to a constant biasing force $F$. The dynamics of this system can be described by the following Langevin equation
\begin{equation}
	\label{model}
	M\ddot{x} + \Gamma\dot{x} = -U'(x) + F + \sqrt{2\Gamma k_B T}\,\xi(t), 
\end{equation}
where the dot and the prime denote differentiation with respect to the time $t$ and the particle coordinate $x$, respectively. The parameter $\Gamma$ is the friction coefficient and $k_B$ is the Boltzmann constant. The spatially periodic potential $U(x)$ is assumed to be in one of the simplest forms, namely
\begin{equation}
	\label{potential}
	U(x) = -\Delta U \sin{\left(\frac{2\pi}{L}x\right)}
\end{equation}
where $\Delta U$ is half of the barrier height and $L$ is the spatial period. Thermal fluctuations due to the coupling of the particle with the thermal bath of temperature $T$ are modeled by $\delta$-correlated Gaussian white noise $\xi(t)$ of zero mean and unit intensity, i.e.,
\begin{equation}
	\langle \xi(t) \rangle = 0, \quad \langle \xi(t)\xi(s) \rangle = \delta(t-s).
\end{equation}
The noise intensity factor $2\Gamma k_B T$ in Eq. (\ref{model}) follows from the fluctuation-dissipation theorem \cite{kubo1966} which ensures that the equilibrium counterpart of the system given by Eq. (\ref{model}) obeys the canonical Gibbs statistics.

In physics only the relations between scales of time, length and energy appearing in physical laws but not their absolute values play a crucial role in determining the progress of observed phenomena. Therefore as the first step we transform Eq. (\ref{model}) into its dimensionless form. This aim can be achived in several ways depending on the choice of the time scale \cite{spiechowicz2020acta}. Here, we define the dimensionless coordinate $\hat x$ and dimensionless time $\hat t$ in the following manner
\begin{equation}
	\label{scaling}
	\hat x = \frac{2\pi}{L} x, \quad \hat t = \frac{t}{t_1}, \quad 
	t_1 = \frac{L}{2\pi} \sqrt{\frac{M}{\Delta U}}. 
\end{equation}
The characteristic time $t_1$ is the inverse of frequency  of small oscillations in the well of the potential $U(x)$.

Under the above  scaling,  Eq. (\ref{model}) is transformed to the form 
\begin{equation}
	\label{dimless-model}
	\ddot{\hat x} + \gamma \dot{\hat x} = \cos{\hat x} + f + \sqrt{2\gamma \theta}\,\hat{\xi}(\hat t),
\end{equation}
where now the dot denotes differentiation with respect to the dimensionless time $\hat t$. We note that the dimensionless mass is $m = 1$ and the remaining rescaled parameters are 
\begin{equation}
	\gamma = \frac{t_1}{t_2} = \frac{1}{2\pi}\frac{L}{\sqrt{M \Delta U}}\, \Gamma,  \quad f = \frac{1}{2\pi}\frac{L}{\Delta U}\, F. 
\end{equation}
The second characteristic time is $t_2= M/\Gamma$ which for a free Brownian particle defines the velocity relaxation time.
The rescaled temperature $\theta$ is given by the ratio of thermal energy $k_{B} T$ to half of the activation energy the particle needs to overcome the original potential barrier $\Delta U$, i.e., 
\begin{equation}
	\theta = \frac{k_B T}{\Delta U}.
\end{equation}
The dimensionless thermal noise $\hat{\xi} (\hat t)$ assumes the same statistical properties as $\xi(t)$, namely, it is a Gaussian stochastic process with vanishing mean $\langle \hat{\xi}(\hat t) \rangle = 0$ and  the correlation function \mbox{$\langle \hat{\xi}(\hat t) \hat{\xi}(\hat s) \rangle = \delta(\hat t-\hat s)$}. From now on we will use only the dimensionless variables and shall omit the hat in all quantities appearing in the Langevin equation (\ref{dimless-model}).
\begin{figure}[t]
	\centering
	\includegraphics[width=0.9\linewidth]{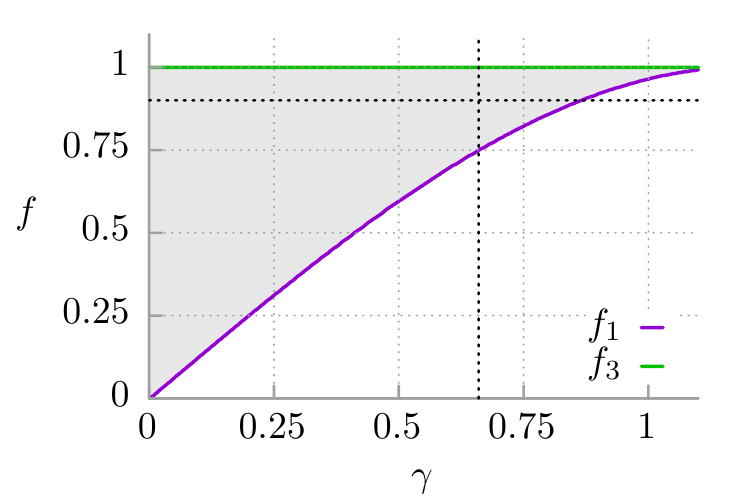}
	\caption{Phase diagram for occurrence of the velocity bistability phenomenon in the deterministic system presented in the dimensionless parameter plane $(\gamma, f)$. Due to this effect in the marked gray area the ballistic diffusion emerges in the deterministic limit of vanishing temperature. The dashed horizontal and vertical line corresponds to $f = 0.9$ and $\gamma = 0.66$, respectively. This regime is analyzed in the paper.}
	\label{risken}
\end{figure}

\section{Deterministic dynamics}
We remind the basic transport properties of the deterministic counterpart of the system, i.e. when $\theta = 0$.  For sufficiently large friction coefficient we can neglect the inertial effects, namely, omit the $\ddot{x} $ term in Eq. (\ref{dimless-model}). Then in the deterministic limit the particle performs a creeping motion. If the tilted potential $\mathcal{U}(x) = -\sin{x} - f x$ exhibits minima, the particle is pinned in the local minimum at large time and its average velocity is ${\mathbb E}[v] = 0$, where 
\begin{equation}
{\mathbb E}[v] = \lim_{t\to\infty} \frac{1}{t} \int_0^t ds \, \dot{x}(s)
\end{equation}
stands for a time averaged velocity for a single trajectory of the system. This solution is termed the locked state. If the bias $f$ is large enough and minima ceases to exist the particle slides down the tilted potential $\mathcal{U}(x)$ and its average velocity is non-zero ${\mathbb E}[v] > 0$. Such a solution is termed the running state. For smaller friction coefficient inertial effects become significant. If the potential $\mathcal{U}(x)$ minima exist in this regime, the locked solution may emerge. Moreover, because of  inertia the  Brownian particle may overcome the potential barriers if the damping is small enough and the running solutions can occur. Such a regime is called bistable and the occurrence of the locked or running state depends on the particle initial state. This effect is well-known since the work of Risken et al \cite{vollmer1983}. To observe average velocity bistability, for a given friction coefficient the constant force must be in range $f_1(\gamma) < f < f_3 = 1$, where $f_1$ is the minimal value for which the running state starts to appear for the deterministic system \cite{risken}. In Fig. \ref{risken} we reproduce the phase diagram for occurrence of the average velocity bistability phenomenon in the dimensionless parameter plane $(\gamma, f)$. This effect is observed only if the bias $f$ lay in the marked gray area.
\section{Overview of diffusion in a washboard  potential}

\subsection{Diffusion quantifiers}
The most fundamental quantity characterizing diffusion process is the mean square deviation (variance) of the particle coordinate $x(t)$, namely \cite{risken}, 
\begin{equation}
	\langle \Delta x^2(t) \rangle = \langle \left[x(t) - \langle x(t) \rangle \right]^2 \rangle = \langle x^2(t) \rangle - \langle x(t) \rangle^2,
\end{equation}
where $\langle \cdot \rangle$ indicates averaging over all thermal noise realizations as well as over initial conditions for the position $x(0)$ and velocity $v(0)=\dot{x}(0)$ of the Brownian particle. In the long time limit the variance typically becomes an increasing function of the elapsing time \cite{metzler2014}
\begin{equation} \label{alpha}
	\langle \Delta x^2(t) \rangle \sim t^{\alpha}.
\end{equation}
The exponent $\alpha$ specifies the diffusion process. Normal diffusion is observed for $\alpha = 1$. The case $0 < \alpha < 1$ is termed as subdiffusion while $\alpha > 1$ describes superdiffusion \cite{metzler2014}. 

One can also define the time dependent "diffusion coefficient" $D(t)$ as \cite{spiechowicz2015pre}
\begin{equation}  
	\label{D(t)}
	D(t) = \frac{\langle \Delta x^2(t) \rangle}{2t}.
\end{equation}
Note that the time-decreasing $D(t)$ corresponds to subdiffusion whereas superdiffusion occurs when $D(t)$ increases. When the exponent $\alpha$ approaches unity then $D(t) = const.$ and we deal with normal diffusion \cite{spiechowicz2015pre}, i.e.
\begin{equation}
	\label{dc}
	D = \lim_{t \to \infty} D(t) < \infty.
\end{equation}
As a result the diffusive behaviour of the system is fully characterized only by specifying both the power exponent $\alpha$ as well as the diffusion coefficient $D(t)$.

\subsection{The riddle of weak noise limit}
Study of various aspects of Brownian motion in a washboard potential has a long history, see e.g. \cite{risken}. This system constitutes a beautiful paradigm of a simple nonlinear system exhibiting  interesting classical and quantum phenomena and still remains  a vibrant topic of current research \cite{guerin2017,zhang2017,kindermann2017,kim2017,fisher2018,cheng2018,dechant2019,goychuk2019,goychuk2020, bialas2020,lopez2020,purello2020}. The reason behind it is its underlying complexity which at the first glance is unnoticeable. Remember that the Fokker-Planck equation for the particle probability distribution $P(x,v,t)$ corresponding to Eq. (\ref{dimless-model}) is a second order partial differential equation of a parabolic type whose parameter space is three dimensional $\{\gamma, f, \theta\}$ and exact solutions are generally unattainable. Nevertheless, some interesting asymptotic and/or limiting cases have been investigated and analytically solved, see e.g.  \cite{risken,lindner2001,reimann2001a,reimann2001b,pavliotis,cheng,derrida}. 

\begin{figure}[t]
	\centering
	\includegraphics[width=0.9\linewidth]{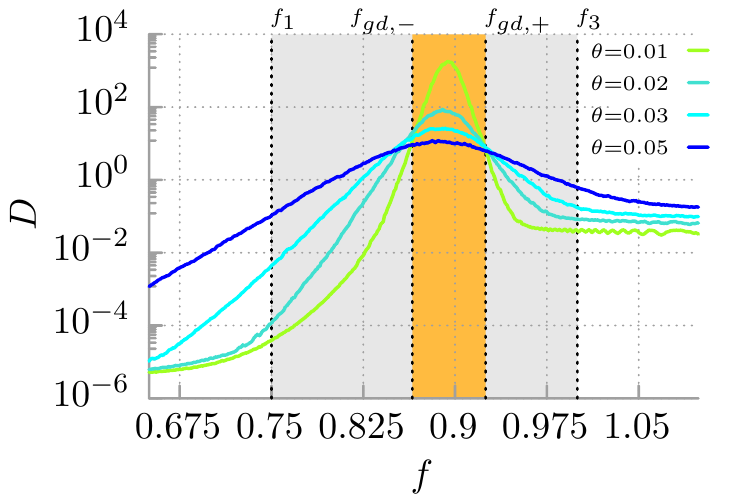}\\
	\caption{The diffusion coefficient $D$ versus the bias $f$ and for different values of temperature $\theta$ of the system. The interval $[f_1,f_3]$ where in the determnistic system bistability of velocity dynamics occurs is shaded by grey color. The critical bias $f$ range $[f_{gd,-},f_{gd,+}]$ for which according to the two-state theory presented in \cite{lindner2016} the diffusion coefficient tends to infinity $D \to \infty$ when temperature vanishes $\theta \to 0$ is marked with an orange region. The dimensionless friction coefficient reads $\gamma = 0.66$.}
	\label{fig2}
\end{figure}

Despite many years of active research on this system one problem still remains unsolved. It is the weak noise limit for diffusion in a tilted periodic potential. 
Amplification of diffusion $D$ by orders of magnitude over the bare diffusion coefficient $D_0 = \theta/\gamma$ of a free particle was first observed  over twenty years ago in \cite{constantini1999}. The authors attributed it to bistability of the velocity dynamics and reported the bell-shaped dependence of the diffusion coefficient on the constant force. Next, in Ref. \cite{lindenberg2005} it has been suggested that the maximal diffusion coefficient $D_{max}$ grows with inverse temperature like a power law 
$D_{max} \sim T^{-3.5}$ and that the force range of diffusion enhancement shrinks to zero when approaching zero temperature $\theta \to 0$. It has been disputed in \cite{marchenko2014} where it was argued that the increase of $D$ follows rather an exponential dependence on the inverse temperature, $D_{max} \sim T^{2/3} \, \mbox{exp}(\epsilon/k_B T)$, where $\epsilon$ is an effective barrier for the bistable velocity dynamics. 
Moreover, the authors reported that when temperature decreases to zero the diffusion coefficient $D$ tends to zero as well (strong damping regime) or it increases (weak damping regime).
Later, in Ref. \cite{lindner2016} a two-state theory was used to determine for all values of the friction coefficient $\gamma$ the range of forces $[f_{gd,-},f_{gd,+}]$, c.f. Fig. \ref{fig2}, in which the diffusion coefficient increases to infinity as temperature decreases to zero, $D_{max} \sim T^{0.22} \, \mbox{exp}(\epsilon/k_B T)$. The authors also indicate that outside of this interval $D$ possesses a pronounced maximum as a function of temperature and it goes exponentially to zero for $\theta \to 0$. The width of the orange region of giant enhancement of diffusion was found to be a non-monotonic function of the friction coefficient $\gamma$ possessing a distinct maximum. The last claim has been discussed later in \cite{marchenko2017,marchenko2019} where the dependence of $D$ on various parameters of the model is discussed and it was suggested that the width of this interval decreases linearly with $\gamma$. Finally, very recently the authors of Ref. \cite{spiechowicz2020pre} followed a different approach and focused on moderate to high temperature regimes to construct a phase diagram for the occurrence of the non-monotonic temperature dependence of the diffusion coefficient. The latter result extends the predictions contained in \cite{lindner2016}, however it barely touches the weak thermal noise limit. Finally, in Ref. \cite{spiechowicz2021arxiv} it is shown that the real-time velocity dynamics in this system is not bistable but rather multistable.

The above brief discussion illustrates that the conundrum of zero temperature limit for diffusion in a tilted periodic potential is still not satisfactory resolved and  there are many mutually contradictory results in the literature. The reason behind it probably lays in the fact that transitions between the velocity states are driven by thermal noise whose intensity reads $\gamma \theta$. Therefore regimes of small damping $\gamma$ and/or temperature $\theta$ require exponentially larger simulation times to sample the state space of the system reliably. Preferably asymptotic analytical methods should be employed, however, these are not yet developed. In this paper we want to debate on the long standing issue of weak thermal noise limit by exploiting the state of the art computer simulations of unprecedented time scale, see below.

\section{Results}
Since we cannot solve analytically the Fokker-Planck equation corresponding to Eq. (\ref{dimless-model}) we had to resort to comprehensive numerical simulations. All numerical calculations have been done by the use of a Compute Unified Device Architecture (CUDA) environment implemented on a modern desktop Graphics Processing Unit (GPU). This proceeding allowed for a speedup of factor of the order $10^3$ times as compared to present day Central Processing Unit (CPU) method \cite{spiechowicz2015cpc}. The quantities characterizing diffusive behaviour of the system were averaged over the ensemble of $10^{5}$ trajectories, each starting with different initial condition $x(0)$ and $v(0)$ distributed uniformly over the intervals $[0, 2\pi]$ and $[-2,2]$, respectively. 

The Langevin equation (\ref{dimless-model}) were integrated using a second order predictor-corrector scheme \cite{platen} with the time step $h = 10^{-2}$. Since we are interested not only in a short time behaviour of the system, but also its asymptotic state, numerical stability is an extremely important problem to obtain reliable results. Hopefully, predictor-corrector algorithm is similar to implicit methods but does not require the solution of an algebraic equation at each step. It offers good numerical stability which it inherits from the implicit counterpart of its corrector.

Each trajectory of the system was associated with its own random number generator, which guaranteed independence of the noise term between different realizations. Initial random number generator seeds were chosen randomly using standard integer random generator available on the host. On GPU we implemented a XOR-shift random number generator that allows for a particularly efficient execution  requiring very small code and state. Its parameters were chosen carefully in order to achieve a period $2^{64} - 1$ which is more than enough for this type of tasks.
\begin{figure}[t]
	\centering
	\includegraphics[width=0.9\linewidth]{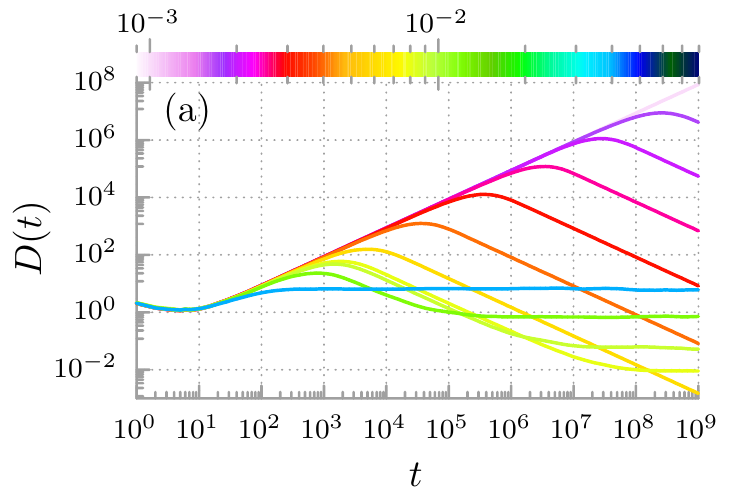}\\
	\includegraphics[width=0.9\linewidth]{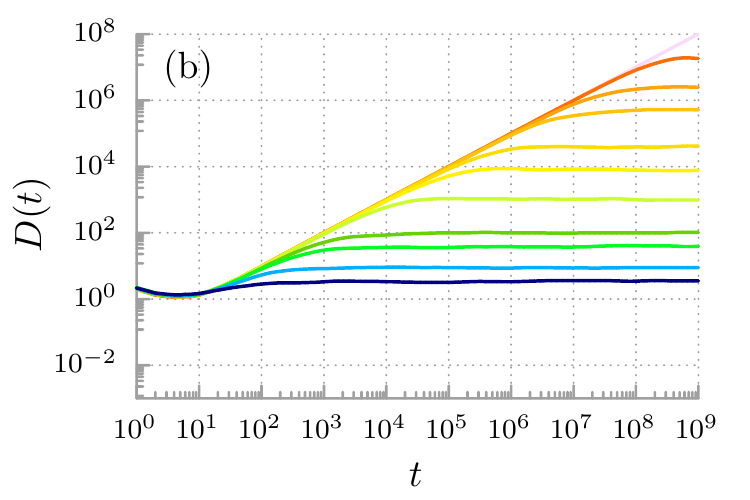}\\
	\includegraphics[width=0.9\linewidth]{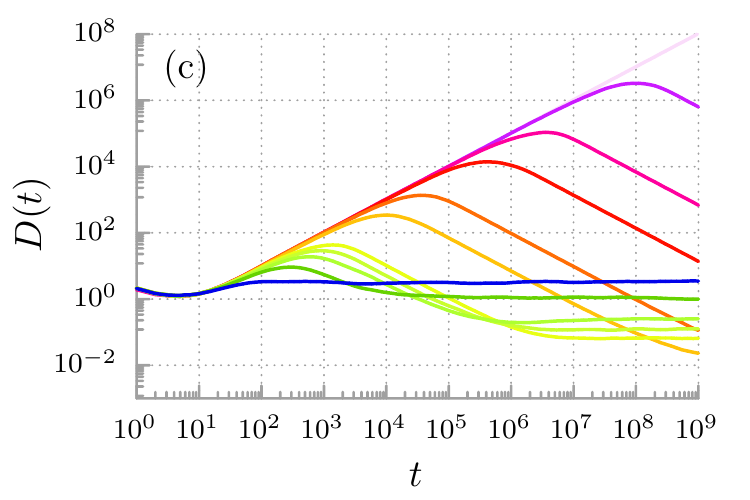}
	\caption{Trajectory of the diffusion coefficient $D(t)$ depicted for different values (color-coded) of the dimensionless temperature $\theta$ of the system. Panel (a): subcritical force \mbox{$f = 0.85 \in [f_1,f_{gd,-}]$}; (b): critical bias $f = 0.9 \in [f_{gd,-},f_{gd,+}]$; (c): supercritical force $f = 0.94 \in [f_{gd,+},f_3]$, c.f. Fig. \ref{fig2}. The friction coefficient reads $\gamma = 0.66$.}
	\label{fig4}
\end{figure}

\subsection{Everlasting anomalous diffusion}
The deterministic counterpart of the setup exhibits the bistable velocity dynamics if for a given friction $\gamma$ the bias is in the range $f_1(\gamma) < f < f_3 = 1$. When $f < f_1(\gamma)$ only the locked solution emerges while for $f > f_3$ the running state occurs exclusively. Therefore for these regions there exists only one class of trajectories $\langle x(t) \rangle \sim {\mathbb E}[v] \,t$ with ${\mathbb E}[v] = 0$ 
or ${\mathbb E}[v] > 0$, respectively. Consequently, there is no spread of trajectories and the diffusion coefficient must vanish outside the gray region in Fig. \ref{risken}, i.e. $D = 0$ when $\theta = 0$ and likewise $D \to 0$ if $\theta \to 0$. 
In the bistable region (inside the gray area) the diffusion is ballistic in the deterministic regime. The contribution comes from the spread between the locked and running trajectories and formally $D = \infty$ in Eq. (\ref{dc}). If temperature is non-zero, the diffusion coefficient $D$ is non-zero and finite in all three intervals of the bias $f < f_1$, $f \in [f_1, f_3]$ and $f>f_3$. In the reminder part of the paper we limit our study to the interval $[f_1,f_3]$, where the bistability of the velocity dynamics occurs and for which there is no consensus on the weak noise limit. 

In Fig. \ref{fig4} we present the most basic characteristics describing diffusive behaviour of the system, namely the time dependent diffusion coefficient $D(t)$ defined in Eq. (\ref{D(t)}), which surprisingly has been missed in most of the recent studies concerning this issue. It is depicted there for the fixed representative friction $\gamma = 0.66$ 
(cf. Fig. \ref{risken}) and different values of the dimensionless temperature $\theta$ of the system coded via the corresponding color. In panel (a) we show this characteristic for the subcritical bias $f = 0.85 \in [f_1,f_{gd,-}]$; (b) the force taken from the critical region $f = 0.9 \in [f_{gd,-},f_{gd,+}]$; (c) the supercritical bias $f = 0.94 \in [f_{gd,+},f_3]$, c.f. Fig. \ref{fig2}. We note that the time scale of the presented evolution is unprecedented and spans \emph{nine orders of magnitude} of the characteristic unit of time. One can observe there two distinct regimes. The first one is visible in panel (b), i.e. for the bias $f = 0.9$ (see Fig. \ref{risken}) laying in the critical interval $[f_{gd,-},f_{gd,+}]$. In this regime $D(t)$ monotonically tends to its time-independent stationary value $D$. First there is an initial stage $(0, \tau_1)$ of ballistic diffusion when the diffusion coefficient increases linearly (note that the scale is logarithmic for both axis) and next for $t > \tau_1$ normal diffusion is approached. On the other hand, the second regime is observed for the sub and supercritical bias $f$, c.f. panel (a) and (c), where except high temperature $\theta$ the diffusion coefficient $D(t)$ displays non-monotonic relaxation towards its stationary value. In the initial interval $(0, \tau_1)$ the diffusion coefficient $D(t)$ grows with time in the ballistic manner, next  there is an extended window $(\tau_1,\tau_2)$ when it decreases indicating subdiffusion and for $t > \tau_2$ finally it reaches its steady state. For sufficiently high temperature the relaxation pattern of $D(t)$ is monotonic and the same as in the critical bias range $[f_{gd,-},f_{gd,+}]$.
\begin{figure}[t]
	\centering
	\includegraphics[width=0.9\linewidth]{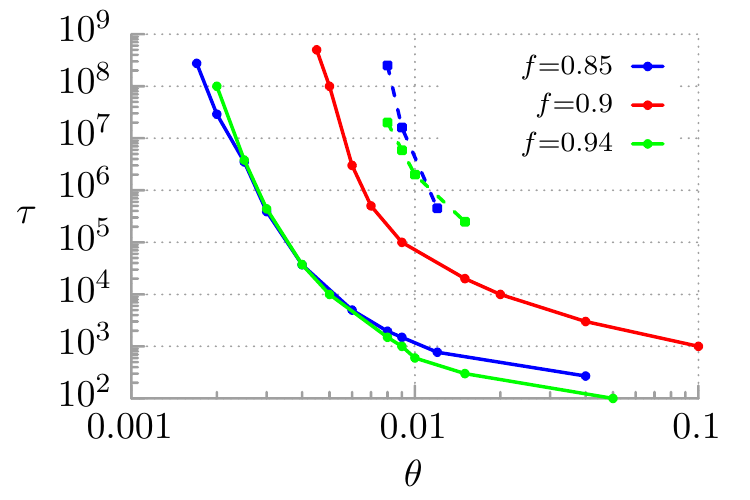}
	\caption{The crossover times $\tau_1$ (solid line) and $\tau_2$ (dashed line) of the ballistic motion and subdiffusion, respectively, are presented versus temperature $\theta$ of the system for the subcritical $f = 0.85$, critical $f = 0.9$ and supercritical $f = 0.94$ bias.}
	\label{fig5}
\end{figure}

Moreover, there are other similarities which are presented in Fig. \ref{fig5} where we depict the influence of temperature $\theta$ on the crossover times $\tau_1$ and $\tau_2$ of the ballistic motion and subdiffusion, respectively. If the temperature decreases $\theta \to 0$ the time-interval of ballistic diffusion rapidly increases and tends to infinity $\tau_1 \to \infty$ regardless of the magnitude of the bias $f$. This picture is consistent with the deterministic limit $\theta = 0$ where in the entire interval $[f_1,f_3]$ the velocity bistability effect occurs which serves as the backbone of the persistent ballistic diffusion $D(t) \to \infty$. 
However, as it is displayed in the panel the crossover time $\tau_1$  depends on the static force $f$. In the critical interval $[f_{gd,-},f_{gd,+}]$ it is much larger than outside of it, e.g. for temperature $\theta = 0.006$ the lifetime of ballistic diffusion reads $\tau_1 \approx 10^7$ for the critical bias $f = 0.9 \in [f_{gd,-},f_{gd,+}]$ and $\tau_1 \approx 10^4$ for both the subcritical $f = 0.85 \in [f_1,f_{gd,-}]$ and supercritical $f = 0.94 \in [f_{gd,+},f_3]$ force. We note that the crossover time $\tau_1$  is  similar for the subcritical and supercritical bias values. The unparalleled time span of our computer simulations allowed as to estimate even the crossover time $\tau_2$ of subdiffusion occuring outside of the critical interval $[f_{gd,-},f_{gd,+}]$. For instance, for temperature $\theta = 0.01$ the lifetime of ballistic diffusion reads $\tau_1 \approx 10^3$ whereas the crossover time of subdiffusion $\tau_2 \approx 10^7$. Therefore the duration of the observed transient anomalous diffusive behaviour is extremely sensitive to temperature variation. Unfortunately, even though our state of the art computational method allowed us to explore exceptionally long time scale of the system evolution we are still far from the truly low temperature regimes. Our results reveal that already for moderate thermal noise intensity $\theta \approx 0.001$ the lifetime of the ballistic diffusion $\tau_1 > 10^9$. However, certainly they are sufficient to draw an inference about the conundrum of weak thermal noise limit for diffusion in a tilted periodic potential. 

One can note that the everlasting anomalous diffusion and the lifetime $\tau_1$ of superdiffusion can be explained in terms of eigenvalues of the Fokker-Planck equation corresponding to the Langevin equation (\ref{model}). In Fig. 11.43(a) in Ref. \cite{risken}, two lowest eigenvalues (say $\lambda_0$ and $\lambda_1$) are depicted in dependence of the bias $f$. The eigenvalues correspond to  characterisic times of the system and determine the rate of  approaching  the stationary state and various stationary expectation values.  The crucial feature is  that inside of  the bistability region, 
 $\lambda_0$ decreases to zero when temperature decreases to zero and  $\lambda_1$ approaches a non-zero value. It means that the characteristic time $\tau_0 =1/\lambda_0 \to \infty$ when $\theta \to 0$. This time can be identified with the lifetime $\tau_1$ of superdiffusion, i.e. $\tau_1 \sim 1/\lambda_0$.  It should be pointed out that in the case of overdamped dynamics, the behavior of eigenvalues as a function of the bias $f$  is different, cf. Fig 3 in Ref.  \cite{lopez2020}. 
 
The results shown in Fig. \ref{fig4} suggest that the time dependent diffusion coefficient $D(t)$ could be described by the following relation 
\begin{equation} \label{E}
	D(t)  \sim  D + E(t), 
\end{equation}
where $D$ is a steady state constant diffusion coefficient defined by Eq. (\ref{dc}).  The function $E(t)$ has the following properties: \\
(i) $E(t) \to 0$ when $t \to \infty$ for any  $\theta \ne 0 $, \\
(ii) its maximum $E_{max} = E(\tau_1) \to \infty$ and  $\tau_1 \to \infty$ when $\theta \to 0$, where $\tau_1$ is the lifetime of ballistic diffusion,\\
Even if for $f \in [f_{gd,-},f_{gd,+}]$, the diffusion coefficient \mbox{$D =D(\theta) \to \infty$} when  $\theta \to 0$ and for  $f \in [f_1,f_{gd,-}]$ or $f \in [f_{gd,+},f_3]$, the diffusion coefficient $D=D(\theta) \to 0$ when $\theta\to 0$, the function $D(t) \to \infty$  as $\theta\to 0$ and in practice it does not make sense to ask about the value of $D$ in the limit $\theta \to 0$ because then $D$ is not well-defined.  

\subsection{Weak ergodicity breaking}
For any non-zero temperature $\theta > 0$ the system (\ref{dimless-model}) is ergodic although the ergodicity is nontrivial since it is driven by a degenerate noise \cite{cheng}. At very low temperature the whole phase space is still accessible because of thermally activated escape events connecting coexisting deterministic disjoint attractors, however, the time after it is fully sampled is extremely long. If temperature tends to zero $\theta  \propto T \to 0$ the crossover time of superdiffusion monotonically increases to infinity $\tau_1 \to \infty$ and ergodicity is broken. In experiments there is not much difference whether the system is non-ergodic or ergodic but exhibiting an unusually slow relaxation towards its steady state. The latter situation occurring here is often captured as weak ergodicity breaking \cite{spiechowicz2016scirep}. It can be characterized by the Deborah number $De$ \cite{reiner1964}
\begin{equation}
	De = \frac{\tau}{\mathcal{T}},
\end{equation}
which is a ratio of a relaxation time $\tau$ of a given observable and the time of observation $\mathcal{T}$. In the case of weak ergodicity breaking it diverges $De \to \infty$. This can happen not only because $\mathcal{T}$ is short but also because $\tau$ is enormously long. In our case the system behaves as weakly nonergodic for the mean square deviation of the particle coordinate when the superdiffusion lifetime $\tau = \tau_1 \sim 1/\lambda_0 $ is sufficiently large. As we demonstrate in Fig. \ref{fig5} this condition is quickly satisfied when temperature $\theta$ goes down to zero.
\begin{figure}[t]
	\centering
	\includegraphics[width=0.9\linewidth]{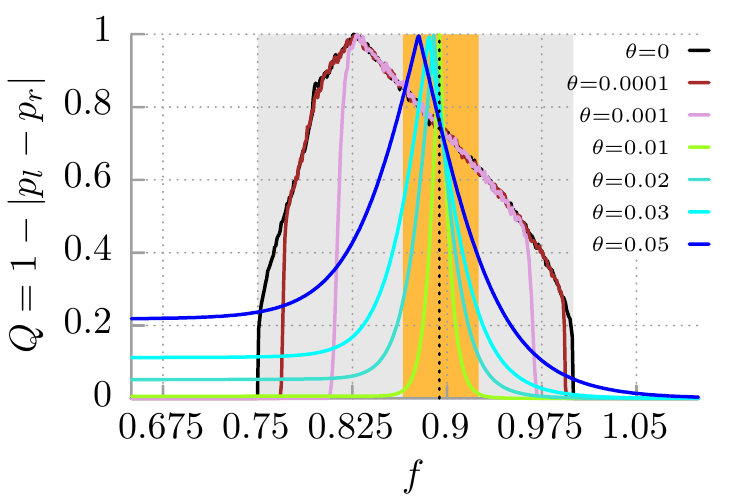}
	\caption{Difference $Q=1 - |p_l - p_r|$ between the stationary probability $p_l$ and $p_r$ of finding the particle in the locked (${\mathbb E}[v] = 0$) and 
	running (${\mathbb E}[v]  > 0$) state, respectively. All presented versus the bias $f$ and for different values of temperature $\theta$ of the system. The interval $[f_1,f_3]$ where in the determnistic system bistability of velocity dynamics occurs is shaded by grey color. The critical bias $f$ range $[f_{gd,-},f_{gd,+}]$ for which according to the two-state theory presented in \cite{lindner2016} the diffusion coefficient tends to infinity $D \to \infty$ when temperature vanishes $\theta \to 0$ is marked with an orange region. The dimensionless friction coefficient reads $\gamma = 0.66$.}
	\label{fig3}
\end{figure}

\subsection{Residence probabilities} 
The properties of the diffusion coefficient $D$ shown in Fig. \ref{fig2} can be reformulated in terms of the stationary probabilities $p_l$ and $p_r$ of finding the particle in the locked (${\mathbb E}[v] = 0$) and running (${\mathbb E}[v]  > 0$) states, respectively. Let us define the following  quantifier $Q = 1 - |p_l - p_r|$ which, roughly speaking, characterizes the difference  between the number of locked and running trajectories.  We note that when $Q \neq 0$ the locked and running solutions coexist. If $p_l=p_r$, i.e. when both trajectories are equiprobable, then $Q$ attains its maximal value $Q=1$. 

For the bistable velocity dynamics and any fixed non-zero temperature $\theta$ there are three contributions to the spread of trajectories of the system around its mean path and consequently to the diffusion coefficient $D$. (i) The first, which is the leading one, is associated with the spread coming from the relative distance between the locked and running solutions. The second (ii) and the third (iii) contribution is related to thermal noise induced spread of trajectories following the locked and running trajectories, respectively. The diffusion coefficient $D$ is maximal when the share of the first one (i) is peaked. It is the case for the equiprobable locked and running trajectories, i.e. when $Q = 1$. We note that in the zero temperature limit $\theta \to 0$ (ii) and (iii) die out and only the first contribution (i) survives. Consequently, the diffusion is ballistic. We exemplify the above observations in Fig. \ref{fig3} for $\theta = 0.01$ by the vertical dashed line pointing at $f = 0.894$,  where the corresponding curve $D(f)$ in Fig. \ref{fig2} is maximal. The resonance-like shape of the quantifier $Q$ provides an elegant yet insightful explanation for the giant diffusion effect which has been overlooked in the literature. Moreover, we pay attention to the fact that the bias $f$ region where the velocity bistability occurs is significantly modified by temperature $\theta$. When $\theta \to 0$ it tends to $[f_1, f_3]$ and not to  $[f_{gd,-},f_{gd,+}]$. We observe that for vanishing thermal noise intensity the number of locked and running trajectories in $[f_1,f_3]$ is of the same order and $|p_l - p_r|$ is relatively small. Consequently, $\langle x (t) \rangle \sim \langle v \rangle t$,  the leading term in the mean square displacement is $\langle \Delta x^2(t) \rangle \sim t^2$  and one observes the persistent ballistic diffusion with $D(t) \to \infty$. Its constancy is guaranteed via the strong ergodicity breaking \cite{spiechowicz2016scirep, spiechowicz2021arxiv}, i.e. there are two mutually inaccessible attractors for the average velocity of the particle in the phase space of the system.

\section{Finite-time measurements} 
\begin{figure}[t]
	\centering
	\includegraphics[width=0.9\linewidth]{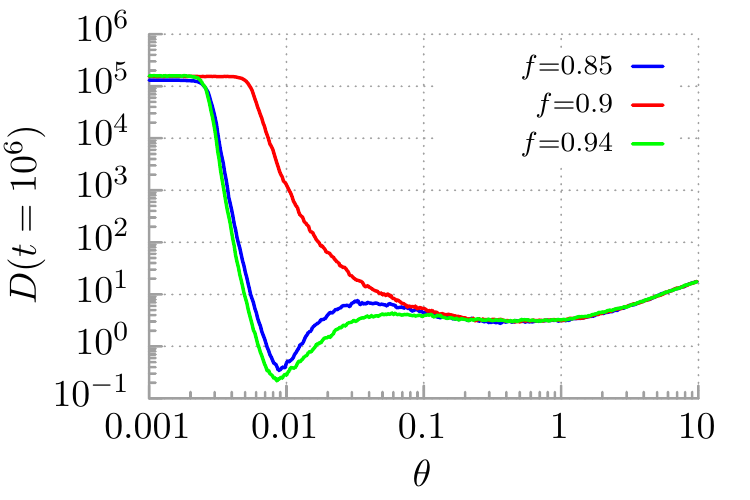}
	\caption{Finite time diffusion coefficient $D(t = t_i = 10^6)$ presented versus temperature $\theta$ of the system for the subcritical, critical and supercritical bias $f$.}
	\label{fig6}
\end{figure}

The time scale of the actual physical experiment is always limited and therefore in the considered case it is useful to analyze the finite time diffusion coefficient $D(t = t_i)$. This characteristics is depicted in Fig. \ref{fig6} for $t_i = 10^6$ as a function of temperature $\theta$ of the system and the subcritical, critical as well as supercritical force $f$. For sufficiently high temperature $\theta$ the Einstein relation $D \propto \theta$ must be recovered as then thermal noise dominates the right hand side of the Eq. (\ref{dimless-model}) and the particle behaves as a free one. Due to the bistability of the velocity dynamics when temperature $\theta$ is lowered we observe a distinctive non-monotonic temperature dependence of the diffusion coefficient $D(t_i)$. We note that the number of diffusion extrema is determined by the magnitude of the bias $f$ acting on the particle. For the critical tilt $f = 0.9 \in [f_{gd,-},f_{gd,+}]$ there is only one minimum whereas for the subcritical $f = 0.85 \in [f_1,f_{gd,-}]$ and supercritical force $f = 0.94 \in [f_{gd,+},f_3]$ both minima and maxima are attainable. However, when we keep lowering temperature eventually the diffusion coefficient $D(t_i)$ saturates on the plateau corresponding to the ballistic diffusion $D(t_i) = D_b(t_i) \sim t_i$. We conclude that in physics limits are often non-commutative. It is the case for diffusion in the tilted periodic potential when the velocity bistability occurs for $f \in [f_1(\gamma),f_3]$. Then (i) if temperature goes down to zero $\theta \to 0$ and time tends to infinity $t \to \infty$ the lifetime of the superdiffusion $\tau_1 \to \infty$ and the diffusion coefficient diverges $D(t) \to \infty$, c.f. Figs. \ref{fig4} and \ref{fig5}; (ii) when temperature vanishes $\theta \to 0$ but time is fixed $t = t_i$ the diffusion saturates on the ballistic front $D(t_i) \to D_b(t_i)$, see Fig. \ref{fig6}; (iii) if temperature is finite $\theta = const.$ but time tends to infinity $t \to \infty$ the diffusion must eventually be normal, i.e. $D(t) = D = const.$, c.f. Fig. \ref{fig4}.

\section{Conclusion}
In this work we revisited the problem of diffusion in a tilted periodic potential to consider the riddle of weak thermal noise limit in this system. It has already been investigated for over twenty years, however it is still not entirely resolved and there are many contradictory results in literature. The reason behind it lies in the fact that in the low temperature regime this system is weakly non-ergodic, i.e. it exhibits an extremely slow relaxation towards its steady state. Therefore asymptotic analytical methods should be employed, however, these are not yet developed.

We exploited the state of the art computer simulations of unprecedented time scale spanning nine orders of magnitude of the characteristic unit of time to study behaviour of the diffusion coefficient $D(t)$ when temperature of the system goes down to zero $\theta \to 0$. We found that for a given damping $\gamma$ if temperature vanishes the diffusion coefficient $D(t)$ in the long time limit (i) is divergent $D(t) \to \infty$ when the constant force is in the range $f_1(\gamma) < f < f_3$, i.e. the velocity bistablity effect occurs; (ii) otherwise it tends to zero $D(t) \to 0$. This result is consistent with the diffusive behaviour observed for the deterministic counterpart of the studied system. In contrast, the statement that for the subcritical $[f_1, f_{gd,-}]$ and supercritical $[f_{gd,+},f_3]$ bias the stationary diffusion coefficient tends to zero $D \to 0$ when temperature vanishes is not compatible with the deterministic dynamics in these regimes. In such a limit the lifetime of ballistic diffusion diverges to infinity and consequently the diffusion coefficient does not reach its stationary value.

We showed that the magnitude of the constant bias $f$ modifies qualitatively the non-monotonic dependence of the finite time diffusion coefficient $D(t_i)$ on temperature $\theta$. In the critical range $[f_{gd,-},f_{gd,+}]$ predicted by a two-state theory it displays only one minimum whereas for the subcritical $[f_1(\gamma), f_{gd,-}]$ and supercritical $[f_{gd,+},f_3]$ area both local minima and maxima are attainable. Last but not least, we demonstrated that for a given temperature of the system the magnitude of the stationary diffusion coefficient $D$ as a function of the bias $f$ is ruled by the relation between the number of the locked and running trajectories. In particular, we identified that the diffusion $D$ is maximal when the locked and running trajectories are equiprobable, i.e. when the difference $Q = 1 - |p_l - p_r|$ attains its maximum $Q = 1$, c.f. Fig. \ref{fig3}.

Summarizing, our numerical results allowed us to draw the inference about conundrum of weak thermal noise limit for diffusion in a tilted periodic potential. These findings for the paradigmatic model of nonequilibrium statistical physics can be applied e.g. to Josephson junctions and cold atoms dwelling in optical lattices.


%
\section*{Acknowledgment}
This work has been supported by the Grant NCN No. 2017/26/D/ST2/00543 (J. S.)

\end{document}